\documentclass[prx,twocolumn,amsmath,amssymb,dvipdfmx]{revtex4} 

\usepackage{color}
\usepackage{bm}
\usepackage{dcolumn}
\usepackage[dvipdfmx]{graphicx}
\usepackage{slashed}
\usepackage{amsmath}\usepackage{accents}
\newcommand{\cin}[1]{{\color{black}#1}}

\begin{document}

\title{
Edge states of a diffusion equation in \cin{one dimension}:\\
Rapid heat conduction to the heat bath
}

\author{Shusei Makino$^1$,  Takahiro Fukui$^1$, Tsuneya Yoshida$^2$, and Yasuhiro Hatsugai$^2$}
\affiliation{$^1$Department of Physics, Ibaraki University, Mito 310-8512, Japan}

\affiliation{$^2$Institute of Physics, University of Tsukuba, 1-1-1 Tennodai, Tsukuba, Ibaraki 305-8571, Japan}

\date{\today}

\begin{abstract}
We propose a one-dimensional (1D) diffusion equation (heat equation) for systems 
in which the diffusion constant (thermal diffusivity) varies alternately with a spatial period $a$.
We solve the time evolution of the field (temperature) profile from a given initial distribution, 
by diagonalising the Hamiltonian, i.e., the Laplacian with alternating diffusion constants,
and expanding the temperature profile by its eigenstates.
We show that there are basically phases with or without edge states. The edge states affect the heat conduction around heat baths.  
In particular, rapid heat transfer to heat baths would be observed in a short time regime, which is estimated to be
$t<10^{-2}$s for $a\sim 10^{-3}$m system and $t< 1$s for $a\sim 10^{-2}$m system
composed of two kinds of familiar metals such as titanium, zirconium and aluminium, gold, etc.
We also discuss the effective lattice model which simplifies the calculation of edge states up to high energy.
It is suggested that these high energy edge states also contribute to very rapid heat conduction in a very short time regime.

\end{abstract}

\pacs{
}

\maketitle

\section{Introduction}

The bulk-edge correspondence  \cite{Hatsugai:1993fk,Hatsugai:1993aa} is recognized as one of fundamental concepts in physics
\cite{Hasan:2010fk,Qi:2011kx}.
The original idea was proposed to answer the question of whether the quantum Hall effect (QHE) is a
bulk or edge-state property \cite{Hatsugai:1993fk,Hatsugai:1993aa},  
but it has now become a way to more broadly characterize topological insulators.
Indeed, it has been playing a central role in the development of topological insulators and superconductors \cite{Kane:2005aa,Qi:2008aa}.
Remarkably, it has been extended to classical  systems such as 
photonic crystals \cin{\cite{Raghu:2008vo,Haldane:2008to,Wang:2009aa,Ozawa:2019us,Kivshar:2019wn}, }
phononic systems \cin{\cite{Prodan:2009vz,Savin:2010wo,Kane:2013aa,Kariyado:2015aa,Susstrunk:2015uo,Chien:2018tk,Yoshida:2019wg,Kivshar:2019wn}, }
electrical circuits \cite{Albert:2015wa,Lee:2018aa,Helbig:2020wq,Yoshida:2020vt},
hydrodynamics \cite{Delplace:2017ui,Sone:2019vz}, and so on.

Recently, the importance of edge states was also pointed out in diffusion phenomena \cite{Yoshida:2021vt}.
This implies that edge states would have a large effect not only on propagating wave systems but also diffusive systems such as
heat conduction.
\cin{In particular, in Ref. \cite{Yoshida:2021vt}, a lattice model was introduced by coarsely discretizing the diffusion equation, and 
their prediction was indeed observed experimentally \cite{hu2021observation,qi2021localized}.
Here, our question is whether the heat conduction  {\it in a continuous medium} that is not necessarily protected by symmetry
is affected by edge states or not.
This attempt is also aimed at understanding  the role of edge states in more mundane phenomena that are not necessarily 
of direct relationship with topology.
}

In this paper,  we examine the one-dimensional (1D) diffusion equation with \cin{position}-dependent diffusion constant, especially paying attention to
the role of edge states. 
In general, one dimensional systems would show anomalous heat conduction due to anomalous diffusion \cite{Li:2003vd}. 
However, 
we assume systems governed by the normal diffusion as well as  the normal Fourier or Fick law, 
and hence, consider the conventional diffusion equation.
Nevertheless, the diffusion equation shows a characteristic behavior when systems allow edge states.

To examine the effect of edge states, we propose 1D diffusion equation with periodic array of two distinct diffusion constants.
\cin{
Most coarse discretization of such an equation with respect to space leads to the Su-Schriffer-Heeger (SSH) 
Hamiltonian \cite{Su:1979aa}, which was already studied in \cite{Yoshida:2021vt}. 
In this case, the relationship between edge states and bulk topology is manifest. In this paper, we investigate the opposite 
limit, a continuous equation for a continuous medium directly, in which the bulk-edge correspondence is rather vague. 
Namely, a continuous medium allows various boundary conditions, and 
the bulk topological invariant protected by inversion symmetry does not necessarily guarantee the edge states on 
boundaries which do not respect inversion symmetry.
Nevertheless,
even without inversion symmetry, we find that there appear many edge states, and 
the diffusion of initial heat distributions given near a boundary is accelerated by edge states in a short time regime.
These edge states are surely related to those that appear for  a boundary condition respecting inversion symmetry.
Thus, we claim that even if boundary conditions breaks symmetry, some of them disappear but others remain persistently. 
}

This paper is organized as follows.
In Sec. \ref{s:diffusion}, we derive a diffusion equation with an $x$-dependent diffusion constant.
We regard it as an imaginary-time Schr\"odinger equation and solve its initial value problem by using the complete set of 
eigenstates.
To this end, we solve the eigenvalue equation of the Hamiltonian first by the Bloch techniques for the bulk system, and next by
the Fourier series expansion under Dirichlet boundary condition for a finite system in contact with the heat baths.
In the latter system, we find 
the appearance of edge states localized at the boundary, which are not present in the bulk system.
Solving the initial value problem, we find 
rapid heat transfer to the heat bath at the boundary, which can be understood as the consequence of edge states.
In Sec. \ref{s:lattice}, we next discretize the diffusion equation and derive an effective lattice model. 
This enables us to obtain the edge states very simply up to very high energies. 
In Sec. \ref{s:summary}, we give summary and discussion including the experimental feasibility.

\section{Diffusion equation in one-dimension}\label{s:diffusion}

We consider generic systems described by  a one-dimensional diffusion equation 
whose diffusion constant takes two values periodically in space.
We assume purely one-dimensional arrays of different materials, or some layered systems stacked in one direction but uniform 
in the directions perpendicular to it.

For a while, we assume  a diffusion equation with a \cin{position}-dependent diffusion constant $D(x)$ generically. 
Let $\phi(t,x)$ be the local field at time $t$.
We assume the Fick law 
\begin{alignat}1
j(t,x)=-D(x)\partial_x \phi(t,x),
\end{alignat} 
where $j(t,x)$ stands for the current density.
Then, the continuity equation reads
\begin{alignat}1
\partial_t \phi(t,x)+\partial_x j(t,x)=0.
\end{alignat}
These equations lead to the following diffusion equation,
\begin{alignat}1
\partial_t\phi(t,x) -\partial_x D(x)\partial_x \phi(t,x)=0.
\label{DifEqu}
\end{alignat}
In the following, we regard Eq. (\ref{DifEqu}) as an imaginary-time Schr\"odinger equation 
$\partial_t\psi+H\psi=0$ and consider the eigenstates of the Hamiltonian operator
\begin{alignat}1
H\phi(x)\equiv-\partial_x D(x)\partial_x \phi(x)=
\varepsilon \phi(x).
\label{DifHam}
\end{alignat}
Note that the Hamiltonian $H$ in Eq. (\ref{DifHam}) is Hermitian.

In what follows, we are mainly interested in the heat conduction.
Then, to be precise, it is necessary to take into account the $x$-dependence of two constants 
characterizing the materials, i.e.,  the heat capacity and thermal conductivity.
As a result, the Fourier law with a $x$-dependent thermal conductivity 
and the continuity equation with a $x$-dependent heat capacity 
lead to more complicated equation than Eq. (\ref{DifHam}).
In particular, the Hamiltonian becomes non-Hermitian in general. 
However, for the sake of simplicity in this paper, we consider only thermal diffusivity $D(x)$ as a $x$-dependent parameter,
ignoring the derivative of  the heat capacity.
Even with this simpler equation, it is possible to analyze edge states, since
they have topological origin, \cin{albeit indirectly,}  as we will see. 
This guarantees the robustness of edge states even if we have resort to 
any approximations, as long as they are small.

\cin{
More basically, 
if one derives  the diffusion equation in inhomogeneous systems microscopically, one starts with the Langevin equation, and 
derives the Fokker-Planck equation. Here, it should be noted that the Langevin equation allows several interpretations
related to the examined process and the nature of the noise, leading to 
different diffusion equations \cite{Leibovich:2019wo,Santos:2020tv}. 
Among them, It\^o and Stratonovich types yield non-hermitian Hamiltonians generically, whereas
in this paper, we adopt the H\"anggi-Klimontovich type \cite{Hanggi:1982va,Klimontovich:1990wt} 
which allows hermitian Hamiltonian (\ref{DifHam}) for simplicity.
We also mention that the Hamiltonian (\ref{DifHam}) can be interpreted as a quantum mechanical kinetic Hamiltonian with
a position-dependent mass \cite{Costa:2020wo}.
}

\subsection{Bulk spectrum}

We now assume $D(x+a)=D(x)$. Then, the Bloch theorem states that the eigenfunction can be written as
\begin{alignat}1
\phi(x)=e^{ikx}u_k(x), 
\label{BloSta}
\end{alignat}
where $-\pi/a<k<\pi/a$ and the Bloch state $u_k(x)$ is periodic, $u_k(x+a)=u_k(x)$.
For 
$u_k(x)$, the eigenvalue equation (\ref{DifHam}) becomes
\begin{alignat}1
-\left[(\partial_xD)(\partial_x+ik)+D(\partial_x+ik)^2\right]u_k=\varepsilon_k u_k.
\label{EigVal}
\end{alignat}
Now, let us expand the periodic functions $D(x)$ and $u_k(x)$ in the Fourier series,
\begin{alignat}1
D(x)=\sum_n e^{ik_nx}d_n, \quad
u_k(x)=\sum_n e^{ik_nx}u_{k,n}, 
\label{DifCon}
\end{alignat}
with $k_n=2\pi n/a$ $(n=0,\pm1,\pm2,\cdots)$.
Substituting these into Eq. (\ref{EigVal}), we have
\begin{alignat}1
\sum_m{\cal H}_{nm}u_m=\varepsilon u_n,
\end{alignat}
where the $k$-dependence of $u_{k,n}$ has been suppressed, and the Hamiltonian is given by
\begin{alignat}1
{\cal H}_{nm}\equiv
\left[k_{n-m}(k+k_m)+(k+k_m)^2\right]d_{n-m}.
\label{ConBulHam}
\end{alignat}
Thus, the Fourier coefficient $d_n$ determines the diffusion phenomena in the present system.

\begin{figure}[htb]
\begin{center}
\begin{tabular}{c}
\includegraphics[width=.7\linewidth]{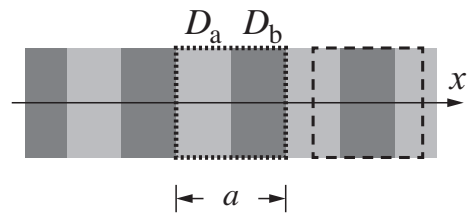}
\end{tabular}
\caption{
Schematic illustration of the system composed of alternating diffusion constants $D_{\rm a}$ and $D_{\rm b}$. 
The dotted line indicates the basic unit cell. The dashed line indicates the alternative unit cell with 
inversion symmetry, which will be used for topological discussions in Sec. \ref{s:topology}.
}
\label{f:ssh_b}
\end{center}
\end{figure}

To be concrete, let us assume
\begin{alignat}1
D(x)=\left\{
\begin{array}{ll}
D_{\rm a}, \qquad\qquad & na<x<\left(n+\frac{1}{2}\right)a,\\
D_{\rm b}, \quad & \left(n+\frac{1}{2}\right)a<x<(n+1)a,\\
\end{array}
\right.
\label{DifConSsh}
\end{alignat}
as illustrated in Fig. \ref{f:ssh_b}.
This can be expressed in the Fourier series such that
\begin{alignat}1
D(x)=\bar D+\frac{\epsilon}{i\pi}\sum_{n={\rm odd}}\frac{1}{n}e^{ik_n x}\equiv
\sum_{n}d_ne^{ik_n x},
\label{DFou}
\end{alignat}
where $\bar D\equiv(D_{\rm a}+D_{\rm b})/2$ and $\epsilon=D_{\rm a}-D_{\rm b}$.

\begin{figure}[htb]
\begin{center}
\begin{tabular}{c}
\includegraphics[width=.99\linewidth]{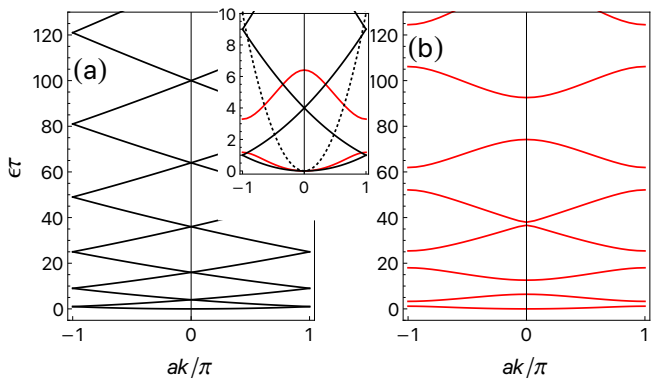}
\end{tabular}
\caption{
Spectra of the model consisting of two kinds of diffusion constants, Eq. (\ref{DifConSsh}).
(a) $(D_{\rm a},D_{\rm b})=(d,d)$, and 
(b) $(D_{\rm a},D_{\rm b})=(10d,d)\mbox{ or }(d,10d)$.
The inset shows the low energy spectra, where
the black and red lines are the same as those in panels (a) and (b). The black dotted line 
is for the uniform case $(D_{\rm a},D_{\rm b})=(10d,10d)$.
}
\label{f:bulk_e}
\end{center}
\end{figure}

In this paper, we assume one typical diffusion constant $d$, and consider systems composed of materials with $d$ and $10d$.
The effect of the ratio $D_{\rm a}/D_{\rm b}$ will  be discussed in Sec. \ref{s:summary}.
In Fig. \ref{f:bulk_e}, we show some spectra of the Hamiltonian (\ref{ConBulHam}).
The Hamiltonian of the uniform system with $D_{\rm a}=D_{\rm b}\equiv d$ ($\epsilon=0$) is a simple Laplacian with 
the dispersion,
\begin{alignat}1
\varepsilon=dk^2\equiv \frac{1}{\tau}\tilde k^2,\quad \tau\equiv \frac{a^2}{\pi^2d},\quad\tilde k\equiv\frac{ak}{\pi},
\end{alignat}
as seen in Fig. \ref{f:bulk_e} (a).  
Once $\epsilon\ne0$ is introduced,
spectrum has a gap at multiples of $k=\pm\pi /a$.
As a result, the lowest band of $D_{\rm a}> D_{\rm b}$ system approaches the band of the uniform system with 
smaller diffusion constant $D_{\rm a}=D_{\rm b}=d$,  not that of uniform systems with $D_{\rm a}=D_{\rm b}=10d$.

So far we have discussed the gapped spectrum of the system with $(D_{\rm a},D_{\rm b})=(10d,d)$.
It may be needless to say that 
the spectrum for the opposite series of the diffusion constant, $(D_{\rm a},D_{\rm b})=(d,10d)$, is completely the same.
Switch between $D_{\rm a}$ and $D_{\rm b}$ is achieved by the transformations 
\begin{itemize}
\item[(r)] reflection: $x\rightarrow-x$ 
\item[(t)] translation: $x\rightarrow x\pm a/2$.
\end{itemize}
These symmetries are for the bulk systems, but approximately apply to the finite systems, as seen in Sec. \ref{s:finite}.

\subsection{Finite system with boundaries}\label{s:finite}

\begin{figure}[htb]
\begin{center}
\begin{tabular}{c}
\includegraphics[width=.8\linewidth]{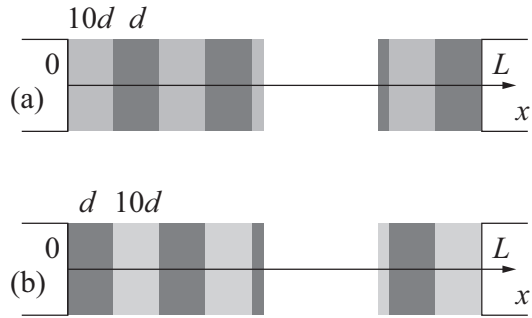}
\end{tabular}
\caption{
Schematic illustration of the finite size system of length $L$. At the boundaries, 
the Dirichlet boundary condition $\phi(0)=\phi(L)=0$ is imposed.
}
\label{f:ssh_e}
\end{center}
\end{figure}

Let us consider the system with length $L$ and impose the Dirichlet boundary condition
which implies,  in the case of heat conduction, a system sandwiched by heat baths at both ends.
See Fig. \ref{f:ssh_e}.
\cin{We assume that the field $\phi(x)$ describes the difference of the temperatures $\phi(x)=T(x)-T_{0}$, where $T(x)$ and $T_{0}$
stand for the temperatures of the material and of the heat baths, respectively. Therefore, we set $\phi(x)=0$ for $x<0,L<x$. 
We also assume that 
within the temperature range of interest, the heat conduction in each layer follows the usual diffusion equation with thermal diffusivity 
$D_{\rm a}$ or $D_{\rm b}$.
}
Then, we can expand $\phi(x)$ in the following
Fourier series
\begin{alignat}1
\phi(x)=\sum_{n>0}\phi_{n}\sin K_nx,
\label{FouSer}
\end{alignat}
where $\phi_{n}$ is real, and  $K_n=\pi n/L$.
Substituting  Eq. (\ref{FouSer}) as well as Eq. (\ref{DifCon}) into  Eq. (\ref{DifHam}), we have
the Hamiltonian 
\begin{alignat}1
{\cal H}_{nm}^{L}=\bar DK_n^2\delta_{nm}\cin{-}\pi_{nm}\frac{\cin{2}\epsilon}{\pi}
\left(\frac{2\pi}{a}S^{(1)}_{nm}K_m-S^{(2)}_{nm}K_m^2\right).
\label{BouHam}
\end{alignat}
Derivations and the definitions of matrix elements are given in Appendix \ref{s:app}.
\cin{In numerical calculations below, we introduce a cutoff $n, m\leq 30$ for the Hamiltonian (\ref{BouHam}).}

\begin{figure}[htb]
\begin{center}
\begin{tabular}{cc}
\multicolumn{2}{c}{\includegraphics[width=.8\linewidth]{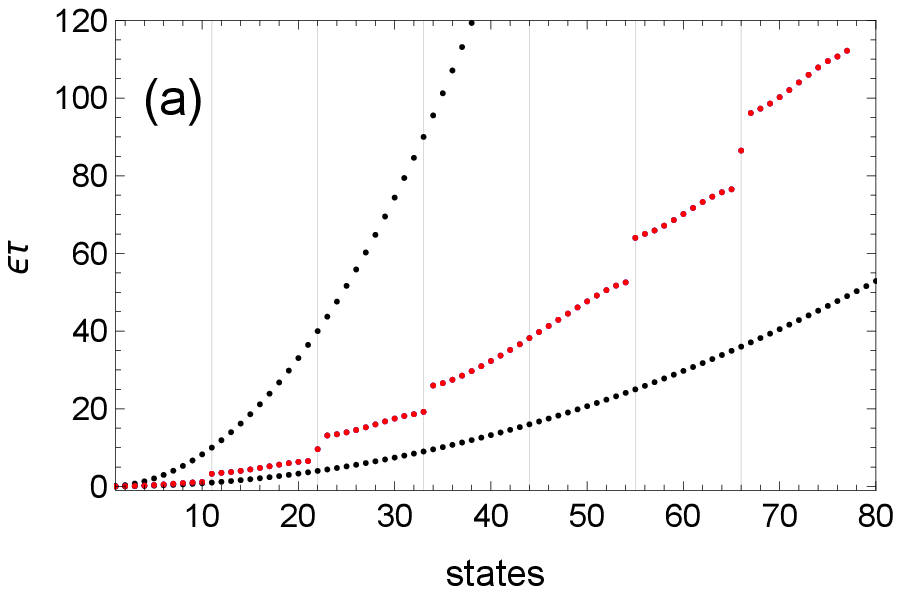}}
\\
\includegraphics[width=.5\linewidth]{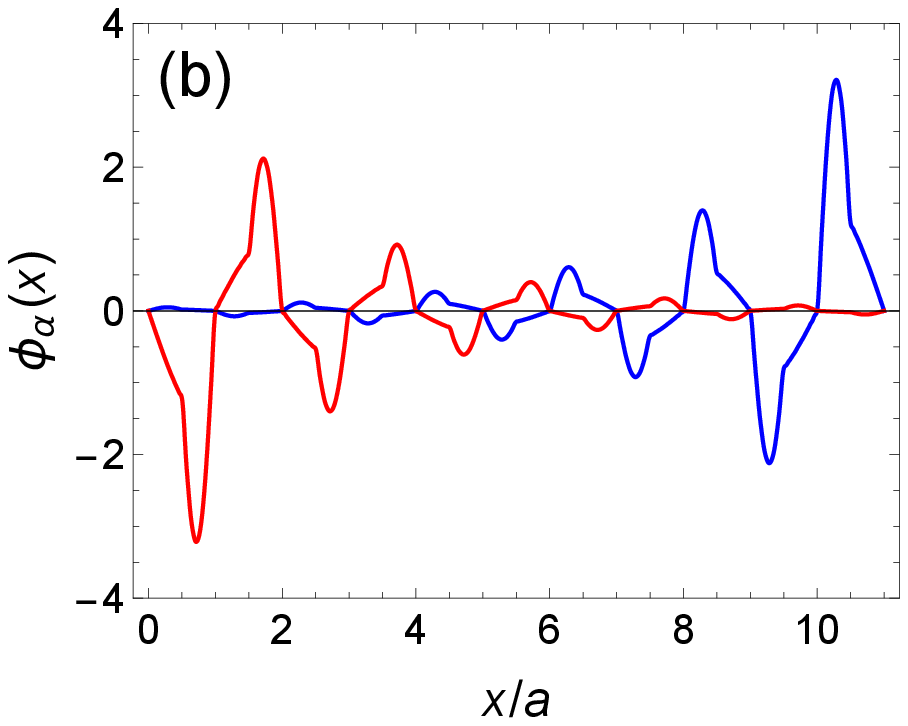}
&
\includegraphics[width=.5\linewidth]{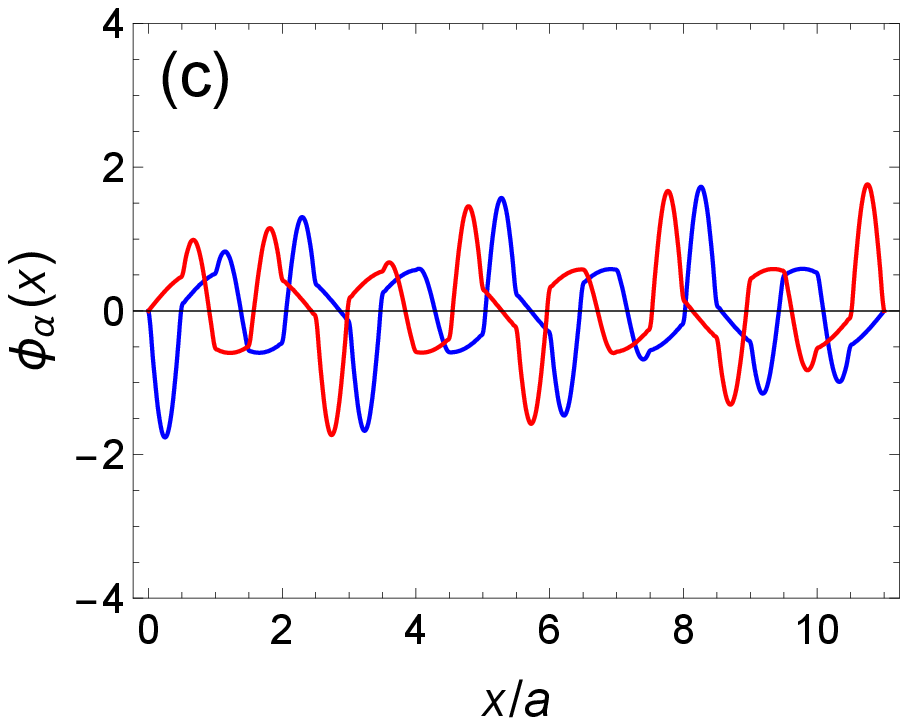}
\end{tabular}
\caption{
(a) Spectra of the finite size system with length $L=11a$. 
The vertical grid lines are guide for the eye,
separating the bulk bands. On some of these lines, the edge states appear.
Red dots are for $(D_{\rm a},D_{\rm b})=(10d,d)$, whereas the black ones are 
 $(D_{\rm a},D_{\rm b})=(d,d)$ for the lower series and $(D_{\rm a},D_{\rm b})=(10d,10d)$ for the upper series.
 Lower two panels show the eigenstate of (b) lowest edge state with $\varepsilon_{\rm edge}\sim3.22/\tau$
 and (c) the 15th extended state with $\varepsilon_{\rm edge}\sim4.32/\tau$.
 Red lines and blue lines are for $(D_{\rm a},D_{\rm b})=(10d,d)$ and $(D_{\rm a},D_{\rm b})=(d,10d)$, respectively.
}
\label{f:edge_f}
\end{center}
\end{figure}

In Fig. \ref{f:edge_f} (a), we show the spectra of the systems with length $L=11a$, solving the eigenvalue equation 
${\cal H}^L\phi_\alpha=\varepsilon_\alpha\phi_\alpha$.
Here, we emphasize that the system with $(D_{\rm a},D_{\rm b})=(10d,d)$ gives exactly the same spectrum, including the edge states,
as the system with $(D_{\rm a},D_{\rm b})=(d,10d)$:
The differences lies in the wave functions.
To see this, 
we show in Fig. \ref{f:edge_f} (b), the eigenfunctions $\phi_\alpha(x)$ ($\alpha=11$),
corresponding to the lowest edge states.
The red  and blue curves are for the systems with $(D_{\rm a},D_{\rm b})=(10d,d)$ and $(D_{\rm a},D_{\rm b})=(d,10d)$, respectively,
corresponding to Fig.~\ref{f:ssh_e}(a) and \ref{f:ssh_e}(b).
In the cases  $D_{\rm a}>D_{\rm b}$ and $D_{\rm a}<D_{\rm b}$, the edge states appear on the left and right ends, respectively.
For reference sake, we also show one example of eigenfunction of the extended state in Fig. \ref{f:edge_f} (c). 
The eigenfunctions of the edge states and extended states have a clear distinction:
While the extended states are related with each other by reflection (r) as well as translation (t), 
the edge states are related only by reflection (t).

\subsection{Time evolution}

Let $\phi_{n\alpha}$ be the normalized $\alpha$-th eigenstate of the Hamiltonian ${\cal H}^L$ in Eq. (\ref{BouHam}). 
Since the Hamiltonian is Hermitian, the eigenfunctions form a complete orthonormal basis,
\begin{alignat}1
\sum_n\phi_{\alpha n}^T\phi_{n\beta}=\delta_{\alpha\beta},\quad \sum_\alpha\phi_{n\alpha}\phi_{\alpha m}^T=\delta_{nm},
\end{alignat}
from which it follows
\begin{alignat}1
\sum_\alpha \phi_\alpha(x)\phi_\alpha(y)=\frac{L}{2}\delta(x-y).
\end{alignat}
Let $\phi_{\rm i}(x)$ be a given initial distribution of $\phi(t=0,x)$. Then, the time evolution of $\phi(t,x)$ is induced by the Hamiltonian 
such that 
\begin{alignat}1
\phi(t,x)&=e^{-Ht}\phi_{\rm i}(x)=\int_0^Ldye^{-Ht}\delta(x-y)\phi_{\rm i}(y)
\nonumber\\
&=\sum_\alpha e^{-\varepsilon_\alpha t}\phi_{\alpha}(x)(\phi_{\alpha},\phi_{\rm i}),
\label{TimEvo}
\end{alignat}
where $\displaystyle(\psi,\chi)=(2/L)\int_0^Ldx\psi(x)\chi(x)$ stands for the inner product of two real functions.

\begin{figure}[htb]
\begin{center}
\begin{tabular}{cc}
\includegraphics[width=.5\linewidth]{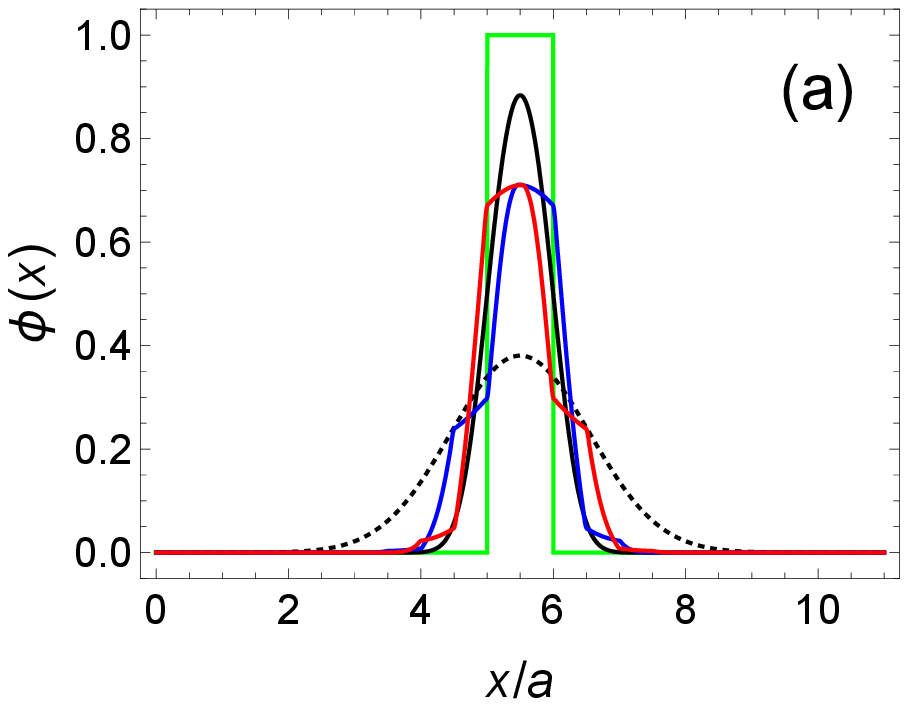}
&
\includegraphics[width=.5\linewidth]{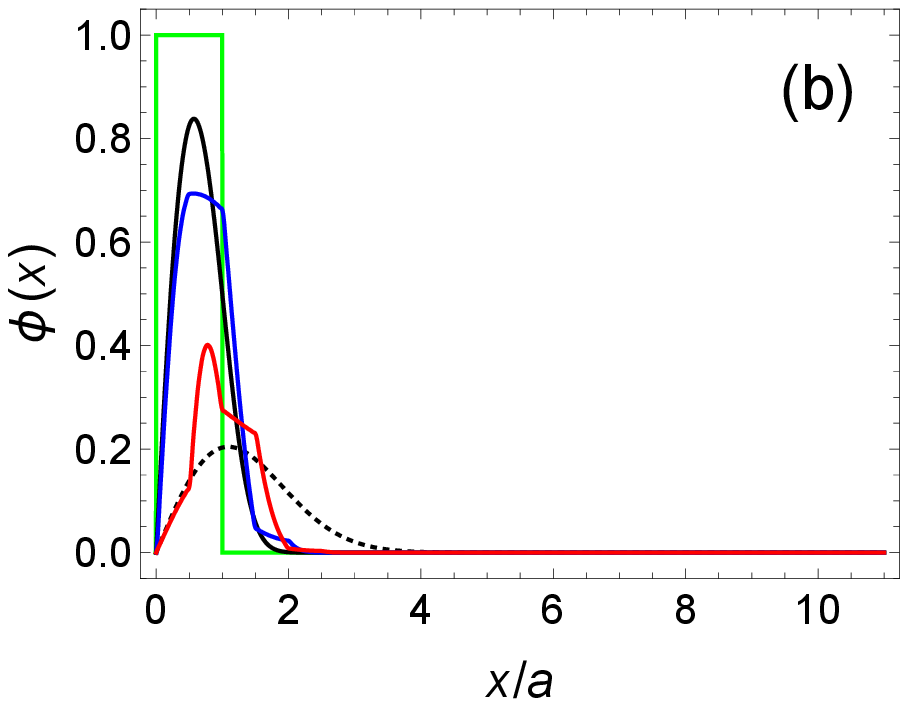}
\end{tabular}
\caption{
The profiles $\phi(t=0.5\tau,x)$ starting from two kinds of initial distributions 
(a) $(\ell_1,\ell_2)=(5,6)$ and (b) $(0,1)$ in Eq. (\ref{IniDis}) denoted by 
green lines.
The system size is $L=11a$.
The red and blue lines are for $(D_{\rm a},D_{\rm b})=(10d,d)$ and $(D_{\rm a},D_{\rm b})=(d,10d)$, whereas black and black-dashed lines
are for $(D_{\rm a},D_{\rm b})=(d,d)$ and $(10d,10d)$, respectively.
}
\label{f:tpro_f}
\end{center}
\end{figure}

In what follows, we give some numerical results for the initial state
\begin{alignat}1
\phi_{\rm i}(x)=
\left\{
\begin{array}{ll}
1\qquad\qquad &\ell_1 a<x<\ell_2 a\\
0& \mbox{(others)}
\end{array}
\right.,
\label{IniDis}
\end{alignat}
where $\ell_j=0,1,2,\cdots$,
implying that a sequence of several unit cells has $\phi=1$ in the background of $\phi=0$.

\subsubsection{Role of edge states}\label{s:role}

We first see that conventional bulk diffusion properties are indeed governed by a smaller diffusion constant $d$ 
rather the larger one $10d$.
To this end, let us give an initial nonzero field at the center of the system, $(\ell_1,\ell_2)=(5,6)$. Then, 
edge states should have nothing to do with the diffusion as far as $t\ll L^2/d\sim10^3\tau$. 
As shown in Fig. \ref{f:tpro_f}  (a), starting from such an initial state,
the profile of $\phi$ at finite $t$ is,  regardless of $D_{\rm a}<D_{\rm b}$ or $D_{\rm a}>D_{\rm b}$, similar
to the uniform system with smaller diffusion constant $(D_{\rm a},D_{\rm b})=(d,d)$ denoted by the black solid curve.
The red and blue curves are slightly shifted to the opposite directions. Physically, it is quite natural, 
while mathematically, it is induced by (t) symmetry of the extended eigenfunctions  
shown in Fig. \ref{f:edge_f} (c).


In contrast, if one gives a finite field $\phi_{\rm i}$ at the boundary unit cell in contact with the heat bath, the diffusion of $\phi$ depends strongly
on $D_{\rm a}<D_{\rm b}$ or $D_{\rm a}>D_{\rm b}$.
The red curve in Fig. \ref{f:tpro_f}  (b) shows  more rapid diffusion than the blue curve.
On one hand, such a difference  seems quite natural, since the initial nonzero field flows directly to the heat bath 
through the diffusion constant next to the bath.
Indeed, in $0<x<a/2$, the red (blue) curve is just on the dashed (solid) curve which is 
the profile with the larger (smaller) diffusion constant.
On the other hand, from the point of view of the formula (\ref{TimEvo}), the difference between the red and blue curves
is attributed to the edge states, 
since in the two systems, (i) the spectra are exactly the same and (ii) bulk states are related with each other by the translation (t), 
whereas only the edge states break the translation (t). 

\begin{figure}[htb]
\begin{center}
\begin{tabular}{cc}
\includegraphics[width=.5\linewidth]{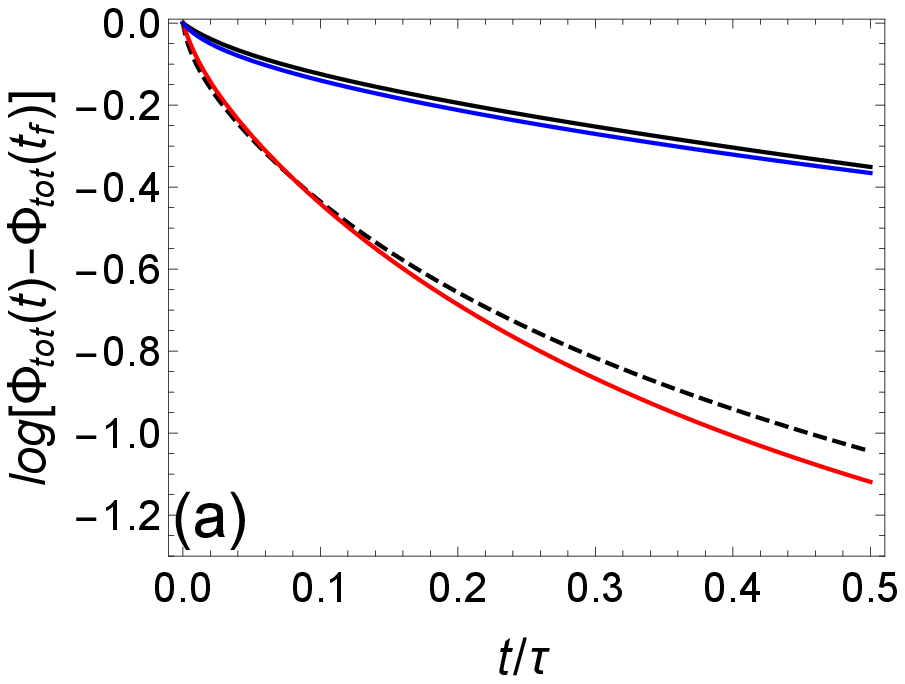}
&
\includegraphics[width=.5\linewidth]{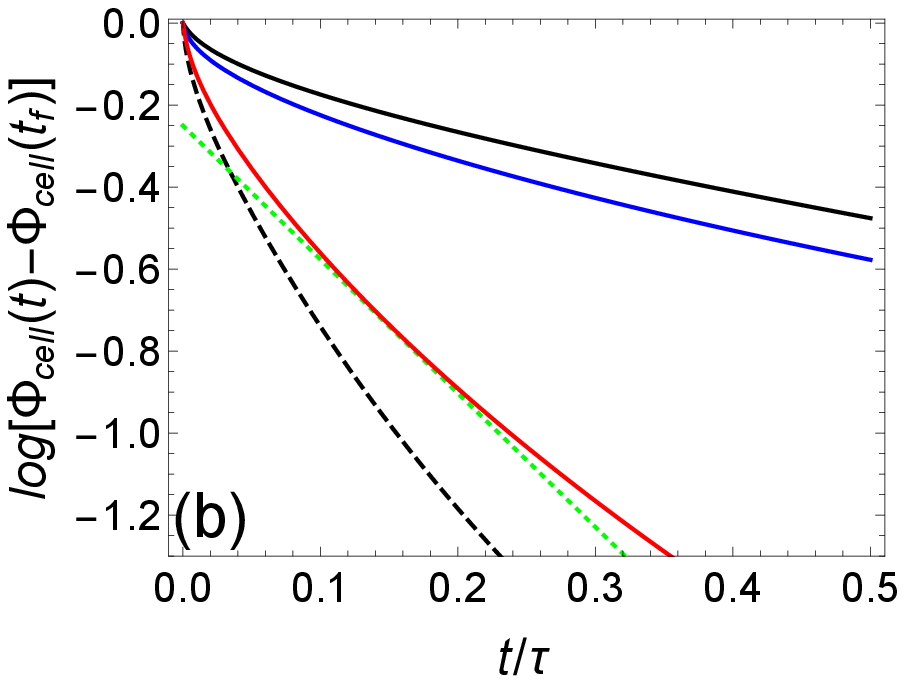}
\end{tabular}
\caption{
(a) Total internal energy and (b) edge internal energy as functions of $t$.
Colors of curves are the same as those in Fig. \ref{f:tpro_f}.
Green dashed line stands for $-\varepsilon_{\rm edge}t+{\rm const}$, where $\varepsilon_{\rm edge}\sim3.3/\tau$ is the energy 
of the lowest edge state, as indicated in Fig. \ref{f:edge_f}.
}
\label{f:Tt_e}
\end{center}
\end{figure}

To clarify this more quantitatively, let us define ``internal energy" by 
\begin{alignat}1
\Phi(t,x_1,x_2)=\int_{x_1}^{x_2}dx'\phi(t,x').
\end{alignat}
In Fig. \ref{f:Tt_e}, we show $\Phi_{\rm tot}(t)\equiv\Phi(t,0,L)$ and $\Phi_{\rm cell}(t)\equiv\Phi(t,0,a)$ 
for the system in Fig. \ref{f:tpro_f} (b).
Figure \ref{f:Tt_e} (a) manifestly shows that the heat transfer to the heat bath is governed by the leftmost diffusion constants.
To see the effects of the edge states, 
we calculate the diffusion from the unit cell at the left edge, shown in Fig. \ref{f:Tt_e} (b).
In the case of $(D_{\rm a},D_{\rm b})=(d,10d)$, there are no low-energy edge states, implying that the diffusion occurs through the bulk extended states only.
Contrary to this, in the case of $(D_{\rm a},D_{\rm b})=(10d,d)$,  quite rapid diffusion is induced at very short time $t/\tau\ll0.1$. 
This is mainly due to high-energy edge states, as will be discussed in Sec. \ref{s:lattice}.
Around $t/\tau\sim0.1$, the lowest energy edge state with the energy $\varepsilon\sim3.2/\tau$ dominates the main 
diffusion which is suggested by the coincidence with the green dashed line.

\subsubsection{Rapid heat conduction}\label{s:rapid}

Since the localization length of the lowest edge state shown in Fig. \ref{f:edge_f} (b) is about $2$ unit cells 
(See also Sec. \ref{s:lattice}), 
the effect of the edge state would be more pronounced 
if we start from a wider distribution of the initial temperature near the boundary.
\begin{figure}[htb]
\begin{center}
\begin{tabular}{cc}
\includegraphics[width=.5\linewidth]{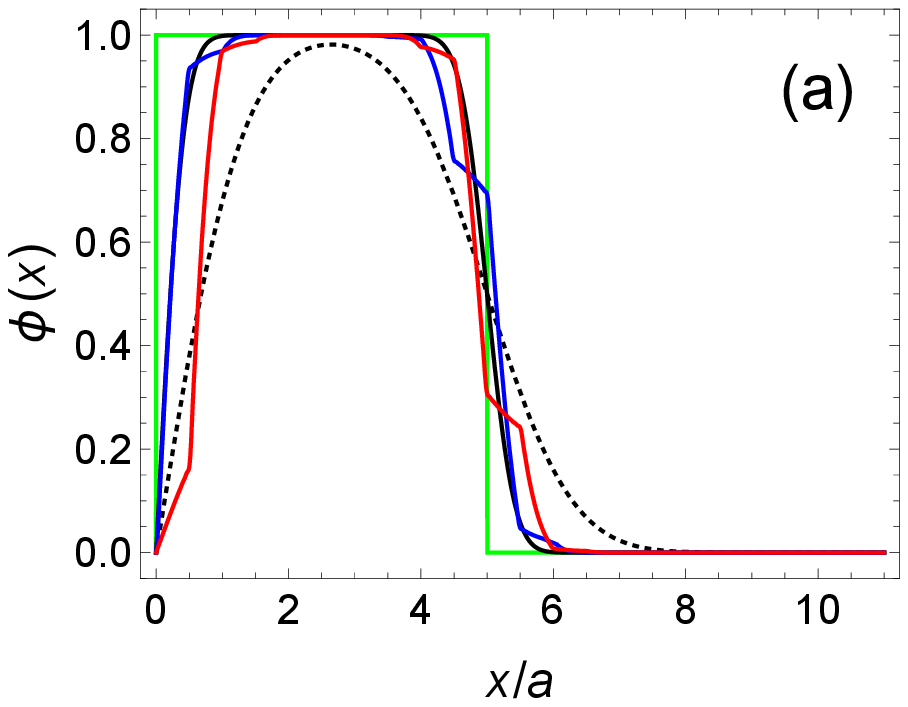}
&
\includegraphics[width=.5\linewidth]{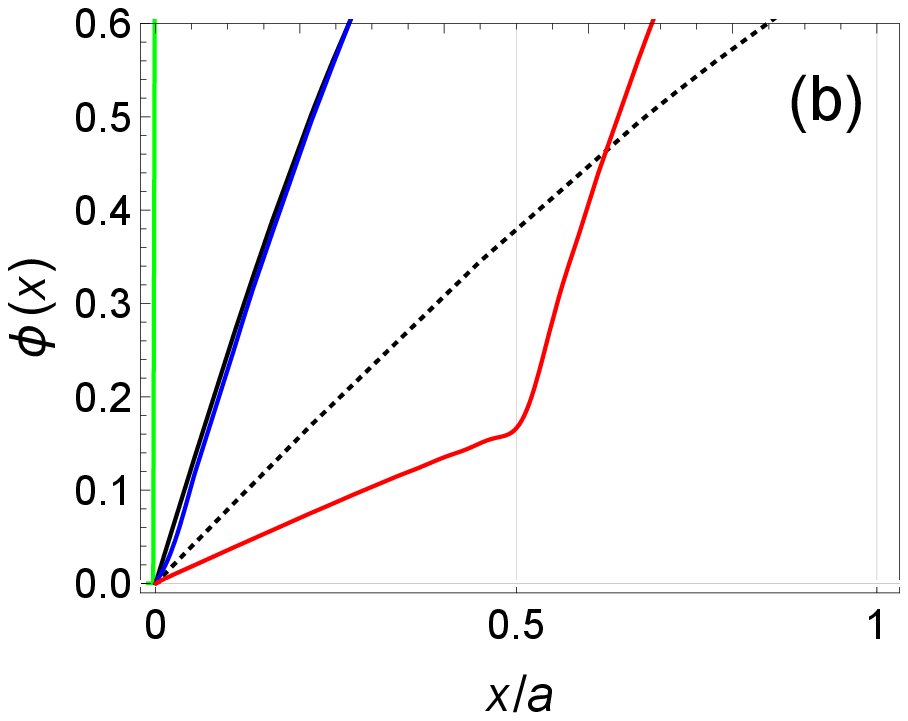}
\end{tabular}
\caption{
(a) Same temperature profile as those in Fig. \ref{f:tpro_f}, but starting from the initial distribution $(\ell_1,\ell_2)=(0,5)$. 
Panel (b) shows the temperature profile of the unit cell at the left end in (a). 
}
\label{f:tpro_f2}
\end{center}
\end{figure}
In Fig. \ref{f:tpro_f2}, the temperature profile starting from the initial distribution $(\ell_1,\ell_2)=(0,5)$ is shown.
Remarkably, rapid heat transfer from the left end to the heat bath is observed for the $(D_{\rm a},D_{\rm b})=(10d,d)$ system.
Indeed, comparing the red curve with the dotted curve, we see that
the temperature at each point of the left end cell with $(D_{\rm a},D_{\rm b})=(10d,d)$ is about half of a uniform system with 
$(D_{\rm a},D_{\rm b})=(10d,10d)$.
Since the temperature profile at $0<x<a/2$ is determined by the balance between the heat escaping into the left heat bath 
and the heat coming in from the right $a/2<x$, the temperature profile of the uniform system with the dashed line 
which is higher than the red curves 
is due to larger heat transfer from the right than that of the alternating system. 
This fact, that heat far from the boundary influences the boundary temperature profile, suggests that 
the temperature profile is determined by the extended states in the uniform system.
On the other hand, the temperature profile in $0<x<a/2$ of the alternating system 
denoted by the red curve is not affected by the heat far from the boundary and is
quite similar to that in Fig. \ref{f:tpro_f} (b), as a result. 
This suggests that such temperature profile is dominated by the edge states.

\section{Effective lattice model}
\label{s:lattice}

So far we have discussed that the edge states modify the short-time heat conduction near the boundaries in contact with heat baths.
Eigenvalues and  eigenstates under the Dirichlet boundary condition have been obtained  using the Fourier series expansion.
In this approach, one cannot judge an eigenstate to be a bulk state or an edge state unless one has a look at the profile of the eigenfunction.
On the other hand,  for tight-binding models, it is very simple to obtain only the edge states separated from the bulk states
\cite{Dwivedi:2016aa,Duncan:2018aa,Kunst:2017aa,Kunst:2018aa,Kunst:2019aa,Kunst:2019ab,Pletyukhov:2020aa,Fukui:2020aa}.
In this section, in order to discuss the edge states of the diffusion equation more simply, we derive an effective tight-binding Hamiltonian by discretizing 
the continuum Hamiltonian in Eq. (\ref{DifHam}).

\begin{figure}[htb]
\begin{center}
\begin{tabular}{c}
\includegraphics[width=.95\linewidth]{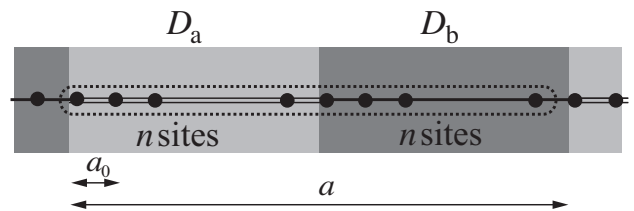}
\end{tabular}
\caption{
Schematic illustration of the discretized system. The dashed-line stands for the unit cell.
}
\label{f:ssh_lat}
\end{center}
\end{figure}

To this end,
let us define the lattice labeled  by $x=ja_0$ ($j=0,\pm1,\cdots$), where $a_0$ is a lattice constant, and introduce 
\begin{alignat}1
\phi_j\equiv \phi(ja_0), \quad D_j\equiv D(ja_0),
\end{alignat}
where $a=2a_0n$. On the lattice, we replace the differential operator in Eq. (\ref{DifHam}) into difference operators defined by
$\partial f_j=(f_{j+1}-f_j)/a_0$ and $\partial^*f_j=(f_j-f_{j-1})/a_0$.
The Hamiltonian in Eq. (\ref{DifHam}) can be discretized such that
\begin{alignat}1
H\phi_j &\equiv -\partial^* D_j\partial \phi_j
\nonumber\\
&=\left[-D_{j-1}\phi_{j-1}+(D_{j-1}+D_{j})\phi_j-D_{j}\phi_{j+1}\right]/a_0^2.
\label{DisHam}
\end{alignat}

It may be convenient to regard $2n$ sites as a unit cell, as illustrated in Fig. \ref{f:ssh_lat}. Namely, we label the sites
such that 
\begin{alignat}1
\phi_j=\phi_{\alpha,J}, \quad j=2nJ+\alpha.
\end{alignat}
Then, the Hamiltonian becomes $2n\times 2n$ matrix operator labelled by $\alpha$.

Since the unit cell can be chosen by an arbitrary $2n$ sequential sites, let us define more generic Hamiltonian 
such that
\begin{widetext}
\begin{alignat}1
H=\frac{1}{a_0^2}\left(
\begin{array}{ccccccc}
D_{2n}+D_{1}&-D_1&&&&&-D_{2n}\delta^*\\
-D_1&D_1+D_2&-D_2&&&&\\
&-D_2&&&&&\\
&&&\ddots&&&\\
&&&&&-D_{2n-2}&\\
&&&&-D_{2n-2}&D_{2n-2}+D_{2n-1}&-D_{2n-1}\\
-D_{2n}\delta&&&&&-D_{2n-1}&D_{2n-1}+D_{2n}
\end{array}
\right),
\label{HamOpeLat}
\end{alignat}
\end{widetext}
where $\delta$ and $\delta^*$ stand for the  forward and backward shift operators defined by
$\delta f_J=f_{J+1}$ and $\delta^* f_J=f_{J-1}$.
The bulk spectrum can be determined by setting $\delta\rightarrow e^{ik}$ and $\delta^*\rightarrow e^{-ik}$.
The wavenumber $k$ of this lattice model, expressed in terms of the wavenumber $k$ of the Bloch state in Eq. (\ref{BloSta}), 
implies $ak$.
Here, we choose $D_1,D_2,\cdots,D_{n}=D_{\rm a}$, and $D_{n+1},\cdots, D_{2n}=D_{\rm b}$ for the effective 
lattice Hamiltonian in Sec.~\ref{s:diffusion}.

\subsection{Edge states}


For tight-binding models, the transfer matrix method is the standard technique to derive the edge states
embedded in the bulk \cite{Hatsugai:1993fk,Hatsugai:1993aa}.
Alternatively, the edge states located at the left end and at the right end can be separately computed 
based on the Hermiticity of the Hamiltonian, as proposed in Ref.~\onlinecite{Fukui:2020aa}.
In this paper, we use the latter method to compute the edge states,
which may be useful for Hamiltonians with large dimensions such as  the discretized Hamiltonian of the present system.
This method is briefly summarized in Appendix \ref{s:edge_ham}.

\subsubsection{Edge state Hamiltonian}

As discussed in Appendix \ref{s:edge_ham},
the edge states localized at the left end are described by the $(2n-1)\times (2n-1)$ Hamiltonian obtained by
neglecting the $2n$th row and column of the Hamiltonian (\ref{HamOpeLat}),
\begin{widetext}
\begin{alignat}1
H_{\rm es}=\frac{1}{a_0^2}\left(
\begin{array}{cccccc}
D_{2n}+D_1&-D_1&&&&\\
-D_1&D_1+D_2&-D_2&&&\\
&-D_2&&&&\\
&&&\ddots&&\\
&&&&&-D_{2n-2}\\
&&&&-D_{2n-2}&D_{2n-2}+D_{2n-1}\\
\end{array}
\right).
\label{EdgStaHam}
\end{alignat}
\end{widetext}
The eigenstates of this Hamiltonian are not necessarily true edge states of the Hamiltonian (\ref{HamOpeLat}): 
We can choose the edge states by requiring that the localization length $1/\kappa$ should be positive,
\begin{alignat}1
e^{-\kappa}=\left|\frac{D_{2n-1}\chi_{2n-1}}{D_{2n}\chi_1}\right|<1.
\label{EdgCon}
\end{alignat}
The edge states at the right end can be also derived in a similar way \cite{Fukui:2020aa}.

\begin{figure}[h]
\begin{center}
\begin{tabular}{c}
\includegraphics[width=.95\linewidth]{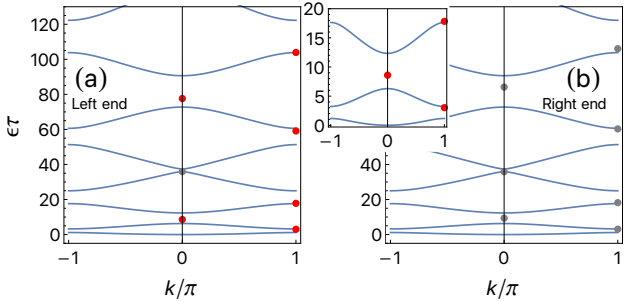}
\end{tabular}
\caption{
(a) Energies of edge states localized at the left end [eigenvalues of the edge state Hamiltonian (\ref{EdgStaHam})] 
with their momenta defined by $k$ in Eq. (\ref{LefCon}) are denoted by colored dots
on the background of the bulk spectra for the model with $(D_{\rm a},D_{\rm b})=(10d,d)$ denoted by lines.
Red (gray) dots show the energies of the edge states (not) satisfying the localization condition (\ref{EdgCon}).
(b) Same as (a) but for the states localized at the right end. There are no states satisfying the localization condition, implying
no edge states at the right end.
Inset shows the states in (a) in the low energy region.
}
\label{f:edge_e}
\end{center}
\end{figure}

In Fig. \ref{f:edge_e}, we plot the eigenvalues of the left and right edge state Hamiltonians 
by colored dots on the background of the bulk bands computed by the Hamiltonian (\ref{HamOpeLat}).
As in the case of the continuum model in Sec. \ref{s:diffusion}, the lattice model with $D_{\rm a}>D_{\rm b}$ 
shows the edge states only at the left end.
Table \ref{t:localization} shows the localization length of the edge states. The lowest edge state has $\sim2$ unit cell localization length, 
implying that the boundary heat conduction is not affected by the initial heat in $x>2$, as indeed observed in Sec. \ref{s:rapid}.

\begin{figure}[h]
\begin{center}
\begin{tabular}{c}
\includegraphics[width=.95\linewidth]{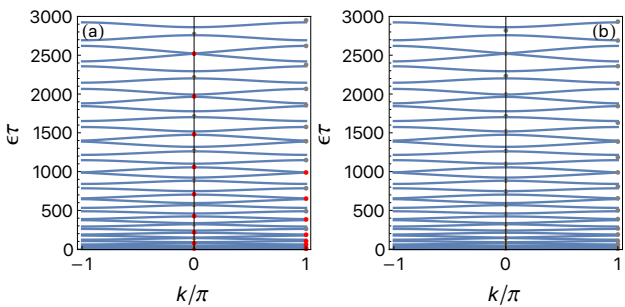}
\end{tabular}
\caption{
Same as Fig. \ref{f:edge_e} up to very high energies. We use $n=100$, so that there can potentially be 199 edge states.
}
\label{f:edge_he}
\end{center}
\end{figure}

In addition to the edge states in Fig. \ref{f:edge_e}, there can be many other edge states up to very high energies.
We extend Fig. \ref{f:edge_e} up to very high energies in Fig. \ref{f:edge_he}. 
The system with $(D_{\rm a},D_{\rm b})=(10d,d)$ is indeed the case, whereas the system with $(D_{\rm a},D_{\rm b})=(d,10d)$ 
never show edge states in all energy regime. Such difference affects the heat conduction at very short time discussed 
in Sec. \ref{s:role}.

\begin{table}
\caption{Energies and localization lengths $1/\kappa$ of the edge states denoted by the red dots in Fig. \ref{f:edge_e}. 
State \# is from the lowest to higher energy.}
\label{t:localization}
\begin{ruledtabular}
\begin{tabular}{lcccccc}
State \#&1&2&3&4&5&6\\
\hline
$\varepsilon$&3.02&8.59&17.83&59.10&77.59& 103.89\\
$1/\kappa$&2.20&0.88&3.23&1.89&0.98&7.36
\end{tabular}
\end{ruledtabular}
\end{table}

\subsubsection{Topological properties}\label{s:topology}

In this section, we argue that  the edge states obtained so far have intimate relationship with the topological property of the bulk system.
For the continuum model, we have chosen the area surrounded by the dotted line in Fig.\ref{f:ssh_b} as the unit cell.
Correspondingly, we have chosen the unit cell of the lattice model to match the continuum model.
In this case, the model has broken inversion symmetry. However, the unit cell in Fig. \ref{f:ssh_b} and corresponding lattice 
model has inversion symmetry. Then, the Berry phase of each band is quantized as $0$ or $\pi$, 
which serves as the topological invariant \cite{Ryu:2002fk}.
It should be noted that bulk spectrum does not depend on the choice of a specific unit cell. The choice of
a unit cell means the choice of the boundaries.

\begin{figure}[htb]
\begin{center}
\begin{tabular}{c}
\includegraphics[width=.95\linewidth]{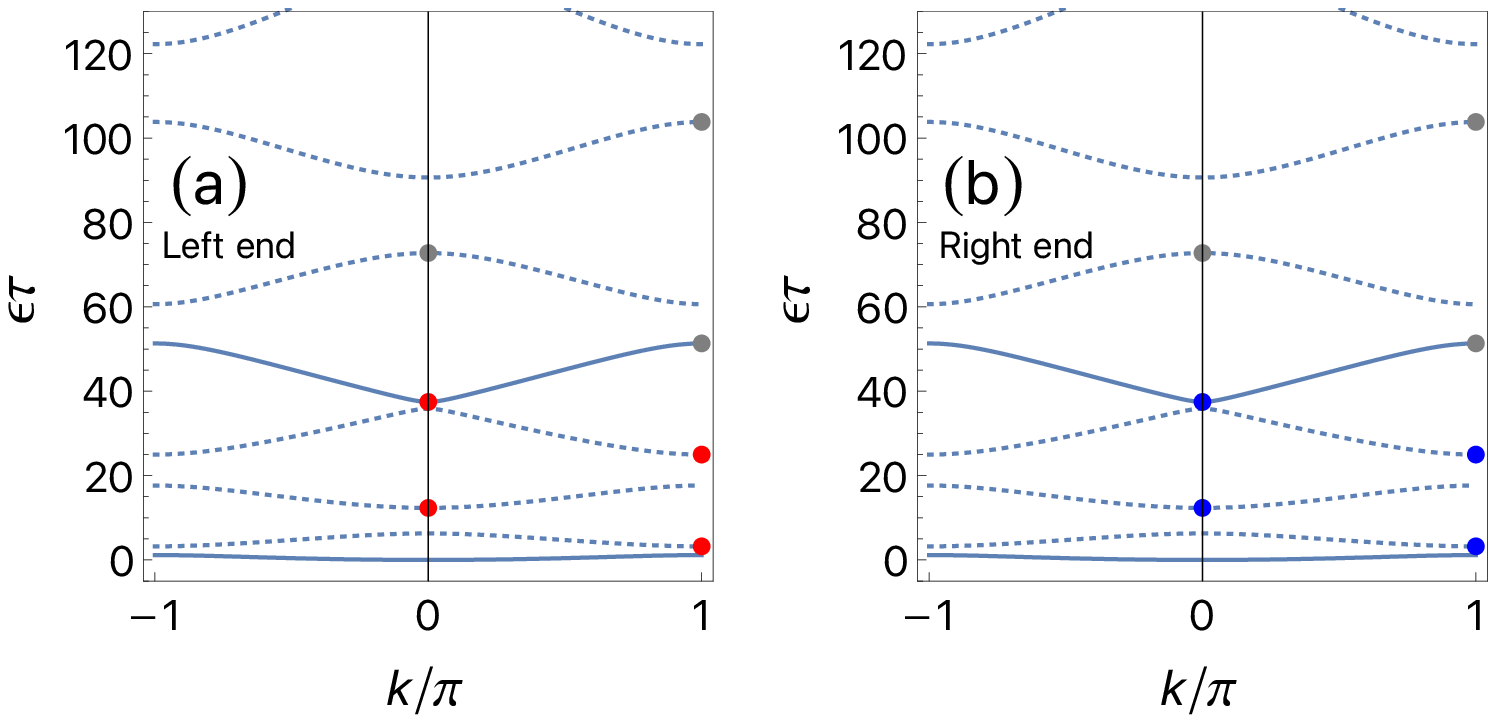}
\\
\includegraphics[width=.95\linewidth]{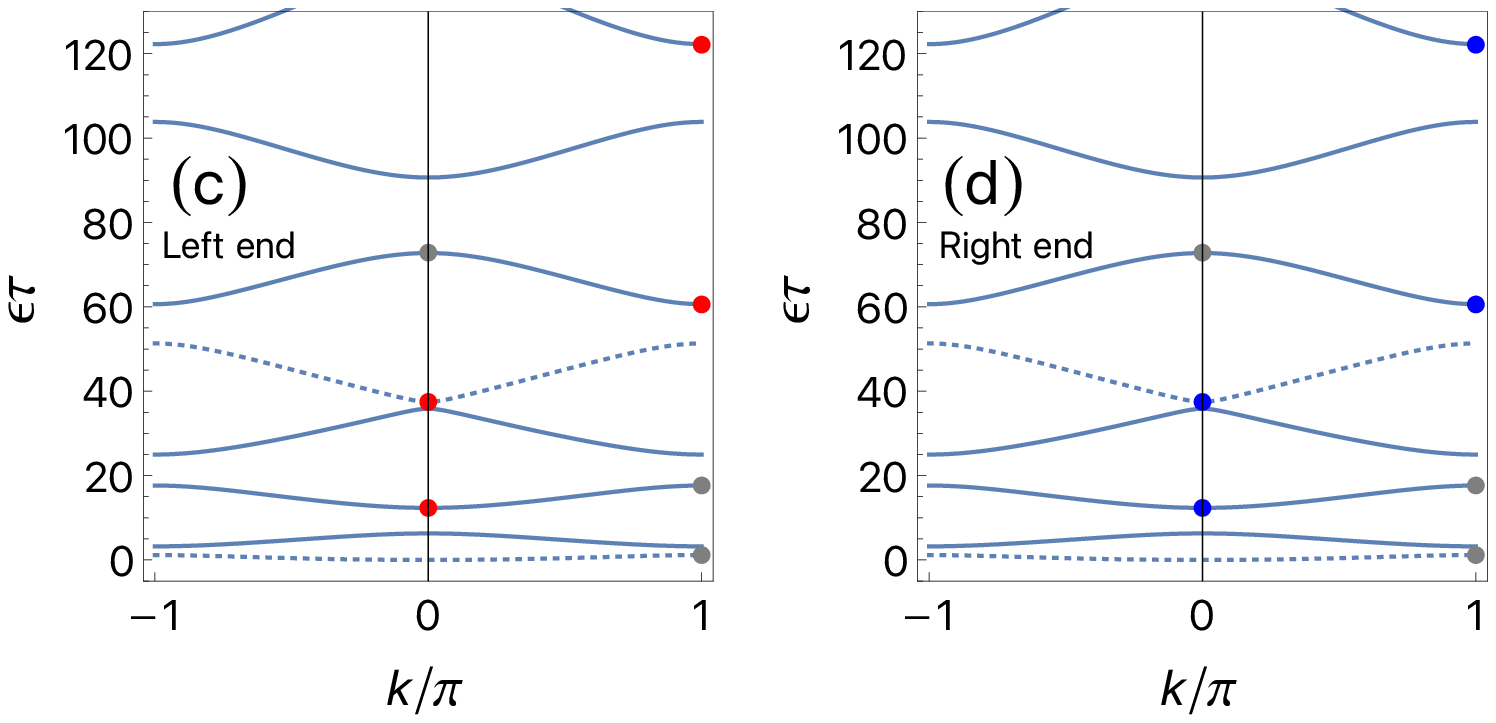}
\end{tabular}
\caption{
Same as Fig. \ref{f:edge_e}, only the unit cell is changed to have inversion symmetry, as illustrated in Fig. \ref{f:ssh_b}. 
The solid lines and dashed lines stand for the bulk bands with Berry phase $\pi$ and $0$, respectively.
The red and blue dots stand for the edge states at the left and right ends, respectively, satisfying the localization condition.
The top two panels (a) are for $(D_{\rm a},D_{\rm b})=(10d,d)$, while the bottom two panels (b) are for
 $(D_{\rm a},D_{\rm b})=(d,10d)$. 
}
\label{f:edge_inv}
\end{center}
\end{figure}

In Fig. \ref{f:edge_inv}, we show the spectra of the lattice model with inversion symmetry.
Each bulk band is distinguished as a solid curve (Berry phase $\pi$) or  a dashed curve (Berry phase $0$). 
With inversion symmetry, the left end and the right end are completely the same, so that the same edge states appear on both ends.
In this system, we can check the bulk-edge correspondence, namely, an edge state appears if the sum of Berry phases 
of the lower bulk bands are $\pi$ modulo $2\pi$.
Comparing Figs. \ref{f:edge_e}, \ref{f:edge_he}, and \ref{f:edge_inv}, 
we conclude that although the edge states in Figs. \ref{f:edge_e} and \ref{f:edge_he} are not directly protected by inversion symmetry,
their origin lies in the symmetry-protected edge states:
Symmetry breaking boundaries make some of them disappear, whereas others remain.

\section{Summary and discussion}\label{s:summary}

We have examined the 1D diffusion equation with the SSH-like alternating diffusion constants.
Since the diffusion equation can be regarded as an imaginary-time Sch\"odinger equation, we have calculated the 
eigenvalues and eigenstates of the Hamiltonian which is the spatial part of the diffusion equation.
We have shown  that the SSH-like structure yields spectral gaps, and if boundaries are introduced, there appear edge states
within the bulk gaps. These edge states have a significant effect on the diffusion process in a short-time regime near boundaries: 
Rapid heat conduction to the heat bath is expected if a boundary allows edge states.

\begin{figure}[htb]
\begin{center}
\begin{tabular}{c}
\includegraphics[width=.99\linewidth]{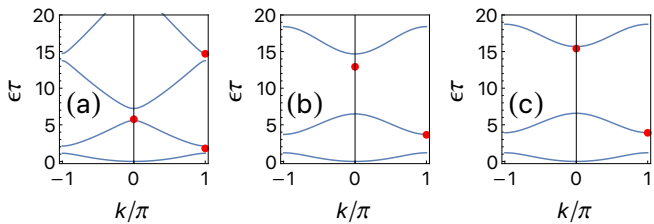}
\end{tabular}
\caption{
Energies of edge states at the left end denoted by red dots. The background liens are bulk bands.
Panels (a), (b), and (c) are the data obtained for $D_{\rm a}/D_{\rm b}=3$, $30$, and $100$, respectively.
The energies and localization lengths of the lowest edge states are (a) 1.78 and 2.17, (b) 3.64 and 3.92, and (c) 3.85 and 6.60.
}
\label{f:edge_s}
\end{center}
\end{figure}

Finally, let us discuss the experimental feasibility. 
The model includes basically two parameters, $D_{\rm a}$ and $D_{\rm b}$. 
We assume $D_{\rm a}>D_{\rm b}\equiv d$, for simplicity. In this case,  the model has edge states at the left end,
which are characterized well by rearranged two parameters $\tau=a^2/(\pi^2d)$ and $D_{\rm a}/D_{\rm b}$.
The key parameter is $\tau=a^2/(\pi^2 d^2)$ 
rather than their ratio $D_{\rm a}/D_{\rm b}$.
To see this, we show the bulk and edge states energies in Fig. \ref{f:edge_s},
changing  the value of $D_{\rm a}$ such that $D_{\rm a}/D_{\rm b}=3,30,100$ (and $10$ in the inset in Fig. \ref{f:edge_e}).
The energies and the localization lengths vary, but qualitative difference is rather small.

Next, let us consider the material-dependence of the parameter $\tau$.
The titanium and zirconium are candidates of materials with smaller thermal diffusivities, $d\sim7\times10^{-6}$ m$^2$/s, 
and hence, fixing the length of the unit cell as $a=10^{-3}$ m, we have $\tau=1.4\times 10^{-2}$~s. 
Combined with materials having larger thermal diffusivities such as carbon ($D_{\rm a}=2.3d$), iron ($2.5d$), lead ($3.4d$),
aluminum ($12d$), and gold ($14d$), rapid heat transfer through edge states discussed in Sec. \ref{s:rapid}
would be observed up to $t\sim 0.5\tau\sim 10^{-2}$ s 
for samples with $a=10^{-3}$ m, and $t\sim 1$ s for samples with $a\sim10^{-2}$ m.

\acknowledgements

This work was supported in part by Grants-in-Aid for Scientific Research Numbers 17K05563, 17H06138,
JP21K13850, and JP20H04627
from the Japan Society for the Promotion of Science.

\appendix
\section{Fourier series expansion under the Dirichlet boundary condition}\label{s:app}
According to the Dirichlet boundary condition (\ref{FouSer}), It may be convenient to rewrite Eq. (\ref{DFou}) with respect to $\sin k_l$,  
\begin{alignat}1
D(x)=\bar D+\frac{2\epsilon}{\pi}\sum_{l={\rm odd}>0}\frac{1}{l}\sin k_l\cin{x}.
\label{DFouSer}
\end{alignat}
Substituting Eq. (\ref{DFouSer}) as well as (\ref{FouSer}) into Eq. (\ref{DifHam}), and using 
\begin{alignat}1
&\int_0^L\sin K_nx \sin K_m x dx=(L/2)\delta_{nm},
\nonumber\\
&\int_0^L\sin K_nx \cos K_m x dx=\frac{2K_n}{K_n^2-K_m^2}\pi_{nm},
\end{alignat}
where $\pi_{nm}=[1-(-)^{n+m}]/2$, we obtain the Hamiltonian (\ref{BouHam}).  The matrix elements associated with 
the couplings with $d_l$ are defined by 
\begin{widetext}
\begin{alignat}1
&S^{(1)}_{nm}=\frac{2}{\pi}\sum_{l={\rm odd}>0}\left[\frac{n+m}{(n+m)^2-(2Nl)^2}
+\frac{n-m}{(n-m)^2-(2Nl)^2}\right],
\nonumber\\
&S^{(2)}_{nm}=\frac{4N}{\pi}\sum_{l={\rm odd}>0}\left[\frac{1}{(n+m)^2-(2Nl)^2}
-\frac{1}{(n-m)^2-(2Nl)^2}\right],
\end{alignat}
\end{widetext}
with $N=L/a$.

\section{Edge state Hamiltonian}\label{s:edge_ham}
The eigenvalue equation of the Hamiltonian (\ref{DisHam}) can be written as
\begin{alignat}1
\left({\cal K}\delta^*+{\cal V}+{\cal K}^\dagger \delta\right)\phi_J=\varepsilon\phi_J,
\label{DisEigEqu}
\end{alignat} 
where $\cal K$ and $\cal K^\dagger$ are matrices proportional to $\delta^*$ and $\delta$, and $\cal V$ is the remaining 
part of the Hamiltonian in Eq. (\ref{HamOpeLat}).
As noted in Sec. \ref{s:diffusion}, the continuum Hamiltonian is Hermitian. 
As a result, the discretized Hamiltonian (\ref{HamOpeLat})
is also Hermitian as far as the system is defined on the infinite line.
However,  if the system has a boundary, the Hamiltonian is not necessarily Hermitian
because of the boundary term due to the summation by parts.
As proposed in Ref.~\onlinecite{Fukui:2020aa}, imposing the Hermiticity condition naturally leads to 
a theoretical framework that allows us to discuss only the edge states in isolation from the bulk states.

Assume that the system has a boundary at $J=1$ and is defined on the semi-infinite line $J\ge1$. 
Then, Hermiticity is guaranteed by
introducing the reference state $\phi_0$ and requiring  
\begin{alignat}1
{\cal K}\phi_0=0.
\label{HerCon}
\end{alignat}
It is readily find that the state $\phi_0^T=(\chi,0)$ satisfies Eq. (\ref{HerCon}), where $\chi$ is a vector with $2n-1$
components. 
Taking $\phi_0$ as a reference state, we can obtain the edge states at the left end by assuming the Bloch-like sates
$\phi_J=\phi_0 e^{iKJ}$ with $K=k+i\kappa$. 
Note that $k$ and $1/\kappa$ are the momentum and the localization length of the edge state, respectively.
The eigenvalue equation then becomes 
\begin{widetext}
\begin{alignat}1
\frac{1}{a_0^2}\left(
\begin{array}{cccccc|c}
D_{2n}+D_{1}&-D_1&&&&&-D_{2n}e^{-iK}\\
-D_1&D_1+D_2&-D_2&&&&\\
&-D_2&&&&&\\
&&&\ddots&&&\\
&&&&&-D_{2n-2}&\\
&&&&-D_{2n-2}&D_{2n-2}+D_{2n-1}&-D_{2n-1}\\
\hline
-D_{2n}e^{iK}&&&&&-D_{2n-1}&D_{2n-1}+D_{2n}
\end{array}
\right)
\left(
\begin{array}{c}
\\
\\
\\
\chi
\\
\\
\\
\hline
0
\\
\end{array}
\right)
=\varepsilon
\left(
\begin{array}{c}
\\
\\
\\
\chi
\\
\\
\\
\hline
0
\\
\end{array}
\right).
\label{EigValEdg}
\end{alignat}
\end{widetext}
Therefore, the edge states are eigenstates of the upper $(2n-1)$ dimensional matrix in Eq. (\ref{EdgStaHam}) with the constraint,
\begin{alignat}1
e^{iK}=e^{ik}e^{-\kappa}=-\frac{D_{2n-1}\chi_{2n-1}}{D_{2n}\chi_1},
\label{LefCon}
\end{alignat}
which follows from the condition to be satisfied by the lowest component in Eq. (\ref{EigValEdg}).
This equation gives the localization length $1/\kappa$ and the momentum $k$ of the edge states by the wave functions.


\end{document}